\documentclass[global,twocolumn]{svjour}
\usepackage{graphicx,cite}
\usepackage[colorlinks=true,citecolor=blue,linkcolor=blue]{hyperref}
\usepackage[section]{placeins}

\newcommand{\subsc}[2]{#1_{\tiny\textnormal{#2}}}
\newcommand{\change}[1]{#1}
\journalname{}
\begin{document}
\title{A tunable low-drift laser stabilized to an atomic reference}
\author{T. Leopold\inst{1} \and L. Schm\"oger\inst{1,2} \and S. Feuchtenbeiner\inst{2} \and C. Grebing\inst{1} \and P. Micke\inst{1,2} \and N. Scharnhorst\inst{1} \and I.D. Leroux\inst{1} \and J.R. Crespo L\'opez-Urrutia\inst{2} \and P.O. Schmidt\inst{1,3} 
}                     
%
%
\institute{ 
Physikalisch-Technische Bundesanstalt, 38116 Braunschweig, Germany \and Max-Planck-Institut f\"ur Kernphysik, 69117 Heidelberg, Germany \and Institut f\"ur Quantenoptik, Leibniz Universit\"at Hannover, 30167 Hannover, Germany}
\date{Received: date / Revised version: date}
%
\maketitle
\begin{abstract}
We present a laser system with a linewidth and long-term frequency stability at the 50~kHz level. It is based on a Ti:Sapphire laser emitting radiation at 882\,nm which is referenced to an atomic transition. For this, the length of an evacuated transfer cavity is stabilized to a reference laser at 780\,nm locked to the $^{85}$Rb D$_2$-line via modulation transfer spectroscopy. 
\change{Gapless frequency tuning of the spectroscopy laser is} realized using the sideband locking technique to the transfer cavity. In this configuration, the linewidth of the spectroscopy laser is derived from the transfer cavity, while the long-term stability is derived from the atomic resonance. Using an optical frequency comb, the frequency stability and linewidth of both lasers are characterized by comparison against an active hydrogen maser frequency standard and an ultra-narrow linewidth laser, respectively. The laser system presented here will be used for spectroscopy of the $1s^{2}2s^{2}2p\ ^{2}P_{1/2} -\ ^{2}P_{3/2}$  transition in sympathetically cooled Ar$^{13+}$ ions at 441\,nm after frequency doubling.
\end{abstract}
\section{Introduction}
\label{sec:intro}
\change{Laser spectroscopy is a powerful tool to gain insight into the structure of atoms and molecules. Spectroscopy of optical transitions in highly charged ions (HCIs) has been proposed for tests of fundamental physics and for superior optical clocks. For example, due to high relativistic contributions to the binding energy of their electrons, some transitions in HCIs are exceptionally sensitive to a change of the fine structure constant \cite{berengut2010enhanced}. Also, small polarizabilities and high internal fields render HCIs insensitive to external perturbations, promising small systematic shifts in atomic clocks based on optical transitions in HCIs \cite{derevianko_highly_2012,dzuba_optical_2015}. However, laser spectroscopy of HCIs is still in its infancy. The highest resolution in laser spectroscopy of HCIs has been demonstrated on the dipole forbidden $^{2}P_{1/2}$ to $^{2}P_{3/2}$ transition in Ar$^{13+}$ with a pulsed dye laser, reaching a resolution of $\sim$\,400\,MHz of this line with an excited state lifetime of 10\,ms \cite{mackel2011laser}. The measurement was limited by the large kinetic energy of the HCIs in the electron beam ion trap. Sympathetic cooling to the mK regime by embedding the Ar$^{13+}$ ions in a laser cooled crystal of Be$^{+}$ ions has been demonstrated in the Cryogenic Paul Trap Experiment (CryPTEx) \cite{schmoger2015coulomb}. The seven orders of magnitude reduction in the temperature of the ions enables Doppler-limited spectroscopy with a resolution on the order of 100\,kHz. Laser spectroscopy of this transition and similar transitions in other HCIs requires a continuous-wave spectroscopy laser that can be adjusted to operate over a large wavelength range and exhibits a linewidth and long-term drift below 100\,kHz as described in this work.}

\change{There are several ways to achieve this goal.} The spectroscopy laser can be locked to a passive cavity \cite{young_visible_1999,kessler_sub-40-mhz-linewidth_2012,hafner_8_2015}, to a stabilized optical frequency comb \cite{benkler_robust_2013,inaba_spectroscopy_2013,rohde_phase-predictable_2014,nicolodi2014spectral,scharnhorst2015high}, or to a broad atomic or molecular transition either directly \cite{wieman_doppler-free_1976,corwin_frequency-stabilized_1998,ye_ultrasensitive_1998,shirley1982modulation,bjorklund_frequency_1983,mccarron2008modulation} or via a transfer cavity \cite{riedle1994stabilization}. In our specific case, the laser will be transported from PTB in Braunschweig, where it is built and calibrated, to the MPIK in Heidelberg, where the HCI spectroscopy will be performed. While stabilization to a passive cavity can in principle provide the required performance goals, a significant engineering effort is required to make it transportable \cite{vogt2011demonstration,chen2014compact,leibrandt2011field,leibrandt2011spherical} \change{ and to eliminate unexpected shifts and non-predictable long-term drifts of its resonance frequencies \cite{dube2009narrow}}. For decades, lasers have been stabilized to atomic references that can in principle provide light with very small long-term drift. 
However, typically two issues have to be overcome: often there are no suitable atomic transitions close to the desired spectroscopy laser frequency, and the linewidth of the transition may be broader than requir ed for the target laser linewidth. 
Both issues can be resolved using a frequency comb referenced to a stable cavity or atomic reference that allows to transfer the stability of the reference to the spectroscopy laser on all relevant time scales \cite{nicolodi2014spectral,scharnhorst2015high}. However, a comb and a suitable reference are required for such an approach, both of which are resource intensive and therefore not generally available in all laboratories. 
\change{For bridging frequency ranges of up to a few THz, an EOM inside a Fabry-Perot cavity can be used to produce a narrow band frequency comb \cite{matthey2015compact}.} As an alternative, a cavity can be used to transfer the stability of a reference laser to the spectroscopy laser \cite{riedle1994stabilization,bohlouli-zanjani_optical_2006,uetake2009frequency,rohde2010diode,albrecht_laser_2012,yin_narrow-linewidth_2015,Bridge:16}. To achieve simultaneous resonance of both lasers with the transfer cavity, at least one of them must be tunable by one free spectral range (FSR). For common cavity lengths the FSR is on the order of 1\,GHz.

Here, we report on a setup to stabilize a spectroscopy laser with a wavelength of 882\,nm via a transfer cavity to a reference laser operating at the Rb D$_2$-line wavelength of 780\,nm. The reference laser is stabilized to the F=3 to F'=4 transition of the $^{85}$Rb D$_2$-line using the modulation transfer spectroscopy technique \cite{shirley1982modulation,mccarron2008modulation}. Frequency drifts of the atomic reference are suppressed by an active temperature stabilization of the Rb vapor cell and magnetic shielding. We use an evacuated cavity to transfer the \change{stability of the reference laser} to the spectroscopy laser, thus avoiding environmental influences that affect the \change{performance of the transfer lock}. The length of the cavity is actively stabilized to the reference laser frequency using a \change{Pound-Drever-Hall (PDH) \cite{black_introduction_2001} sideband locking technique to provide gap-less tuning.} The spectroscopy laser is stabilized to the transfer cavity in a standard PDH configuration.

\change{Using an optical frequency comb that is locked to a long-term-stable H-maser,} we show that the resulting frequency stability of the reference and spectroscopy laser is below 50\,kHz on a time scale of several days. The short term linewidth is measured via a comparison with an ultra-stable laser. Our setup enables us to narrow the linewidth of the spectroscopy laser to well below that of the reference laser, which reduces the requirements for the latter. \change{We operate the laser system at 882\,nm to produce the spectroscopy wavelength of 441\,nm for Ar$^{13+}$ after frequency doubling. However, the Ti:Sapphire laser can be broadly tuned to access optical transitions in other HCIs. The current set of transfer cavity mirrors supports a wide range of wavelengths spanning from $775\dots 890$\,nm. This range can be further extended to other spectroscopy laser wavelengths by adapting the mirror coating.}

\section{Experimental setup and results} \label{sec:setup}

\begin{figure}[htb]
\centering\includegraphics[width=0.45\textwidth]{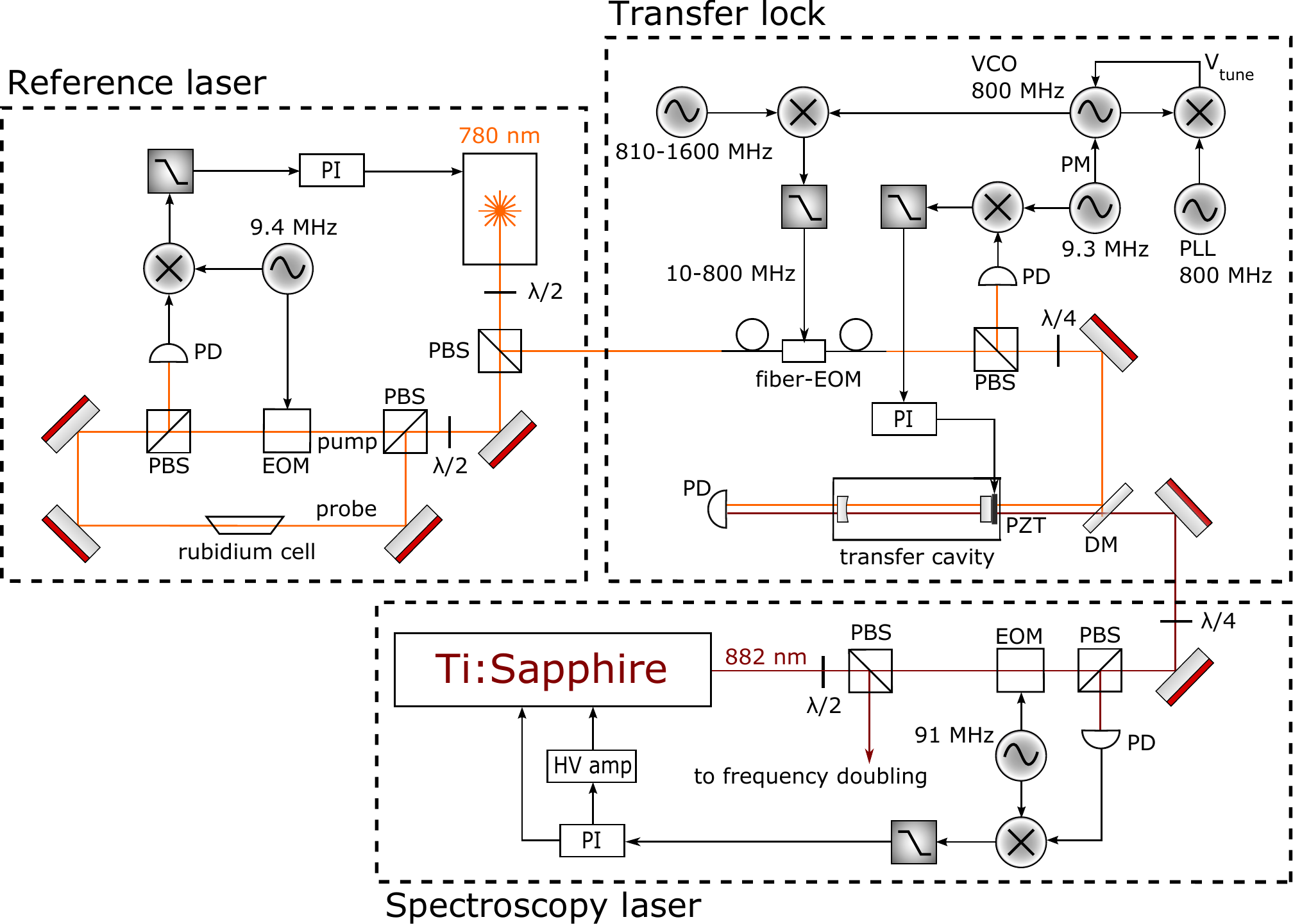}
\caption{Simplified experimental setup showing the reference laser, transfer lock setup and spectroscopy laser. For details about the individual parts of the setup refer to the corresponding sections. PBS: polarizing beam splitter, EOM: electro-optic modulator, PD: photo diode, DM: dichroic mirror, PZT: piezo actuator, PI: proportional-integral controller, VCO: voltage controlled oscillator, PLL: phase-locked loop, HV amp: high-voltage amplifier, PM: phase modulation, $\lambda/2$: half-wave plate, $\lambda/4$: quarter-wave plate, V$_{\textnormal{tune}}$: mixer output/feedback voltage\label{setup}}
\end{figure}

The experimental setup is schematically shown in Fig.\,\ref{setup}. It is divided into three parts: the reference laser, the cavity transfer lock, and the spectroscopy laser, each of which will be discussed in detail in the following.

\subsection{Reference laser} \label{sec:ref}

The reference laser consists of a distributed feedback laser diode (Eagleyard EYP-DFB-0780-00080) mounted in a temperature-stabilized housing (Thorlabs TCLDM9). \change{The low noise laser diode driver (Thorlabs LDC201CU) allows modulating the diode pump current with a bandwidth of 100~kHz for frequency stabilisation.} The free running linewidth of the laser diode is specified to be around 2\,MHz. The output power of $\sim$\,50\,mW is divided into two parts: 5\,mW are used to stabilize the frequency to an atomic resonance by modulation transfer spectroscopy (MTS) \cite{shirley1982modulation,mccarron2008modulation}, the remainder is sent \change{through a fiberized electro-optic modulator to provide the modulation for locking the length of} the transfer cavity.

The optical setup for the frequency stabilization of the reference laser is shown in Fig.\,\ref{setup} in the upper left box. A linearly polarized pump beam is phase modulated at $\subsc{\omega}{MTS}\simeq 9.4$\,MHz. It is aligned to counterpropagate along an unmodulated probe beam in a temperature-stabilized rubidium vapor cell. 
\change{The laser is subject to dispersion in the rubidium vapor which, around resonance, converts the phase modulation of the pump beam to amplitude modulation. Thus, the saturation of the medium becomes} modulated with $\subsc{\omega}{MTS}$. The counterpropagating probe beam experiences a modulated absorption, which is subsequently detected as amplitude modulation on a photo diode. Homodyne mixing of the photo diode signal with $\subsc{\omega}{MTS}$ yields a dispersion-shaped output signal around the atomic resonance. \change{This MTS signal is tailored by a PI controller and fed back as a modulation of the current via the laser diode driver with a typically achieved feedback bandwidth of 7~kHz.}\par 

\begin{figure}[htb]
\centering\includegraphics[width=0.45\textwidth]{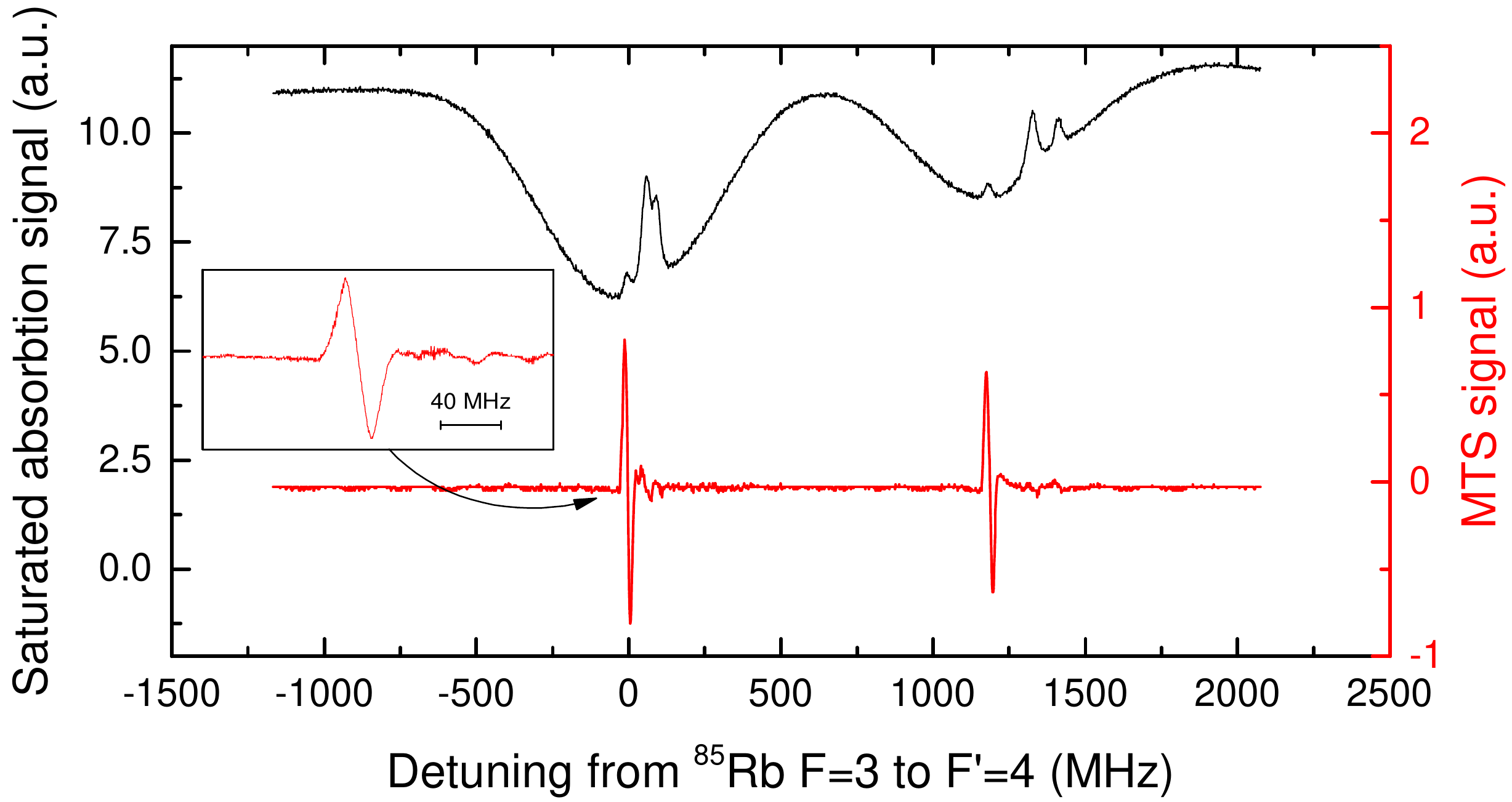}
\caption{Saturated absorption (black) and modulation transfer spectroscopy signal (red). Two Doppler broadened transitions contribute to the D$_2$-line for each isotope. Each consists of three hyperfine transitions. Here the transitions for $^{85}$Rb F=3 to F'=2,3,4 and $^{87}$Rb F=2 to F'=1,2,3 are shown. Only the cycling transitions F=3 to F'=4 ($^{85}$Rb) and F=2 to F'=3 ($^{87}$Rb) produce a strong dispersive signal in the modulation transfer spectroscopy. The inset shows \change{a higher resolution scan across} the $^{85}$Rb transition used to stabilize the reference laser. \label{MTS}}
\end{figure}
In Fig.\,\ref{MTS} the saturated absorption and MTS signals are shown when scanning the laser frequency \change{across} the D2-line\change{s} originating from the F=3 hyperfine ground state in $^{85}$Rb and the F=2 ground state in $^{87}$Rb \cite{steckrubidium85,steckrubidium87}. The advantages of the MTS technique for absolute frequency stabilization can be seen in the figure. Only the closed-cycle transitions F=3 to F'=4 in $^{85}$Rb and F=2 to F'=3 in $^{87}$Rb create a \change{significant} error signal, resulting in a simple identification and reproducibility of the lock point. Furthermore, \change{in contrast to frequency modulation or polarization spectroscopy \cite{mccarron2008modulation} MTS is free of a Doppler background}. 
It has been shown that an absolute frequency accuracy below 1\,kHz can in principle be achieved using MTS \cite{MartinezdeEscobar:15}.\par 
The vapor cell is temperature stabilized to $\sim$\,10\,mK root mean square (RMS) deviation from the set-point. It is magnetically shielded with two insulated layers of mu-metal. A housing around the cell mount with small openings for the laser beam reduces air flow.\par 
The reference laser frequency is locked to the F=3 to F'=4 hyperfine transition in $^{85}$Rb which exhibits the error signal with the steepest slope in our setup. 

\subsection*{Performance of the reference laser} \label{sec:perf_ref}
\change{The reference laser performance and systematic shifts are studied by comparing the laser frequency to a hydrogen maser. For this, part of the laser power is overlapped with the output of a maser-stabilized optical frequency comb and a beat note is observed on a high bandwidth photo diode and recorded by frequency counters. We use two independent counters to validate the frequency data.}

Since the rubidium vapor is enclosed in a fixed volume, any temperature change translates directly into a pressure change. \change{This} not only broadens the transition lines, but also shifts them in energy \cite{PhysRev.124.800}. By changing the set-point of the temperature stabilization while observing the beat frequency over time on an electronic counter, a temperature sensitivity of $\sim$\,100\,kHz/K is measured for the transition \change{used here. The temperature fluctuations of the stabilized cell are $\sim$\,10\,mK,} leading to a frequency instability on the few-kHz level. 
 
The differential magnetic field shift of the hyperfine levels used for laser stabilization is 2.3\,MHz/mT \cite{steckrubidium85}. With typical magnetic field fluctuations on the order of $10^{-2}$\,mT and shielding as described above, magnetic field shifts are negligible on the kHz level.\par 
\change{The error signal of MTS consists of a signal with odd symmetry with respect to the resonance frequency (originating from phase modulation) and a signal with even symmetry (originating from amplitude modulation). Residual amplitude modulation (RAM) at the EOM resonance frequency thus introduces an asymmetry in the error signal \cite{jaatinen1998possible}. The effect of varying laser power on the error signal is a linear scaling of the voltage response. For a symmetric error signal this only affects the slope. For an asymmetric signal, however, the position of the zero-crossing on the frequency axis depends on laser power. RAM-induced asymmetry thus introduces a laser power dependence of the lock point. 
}

\begin{figure}[htb]
\centering\includegraphics[width=0.45\textwidth]{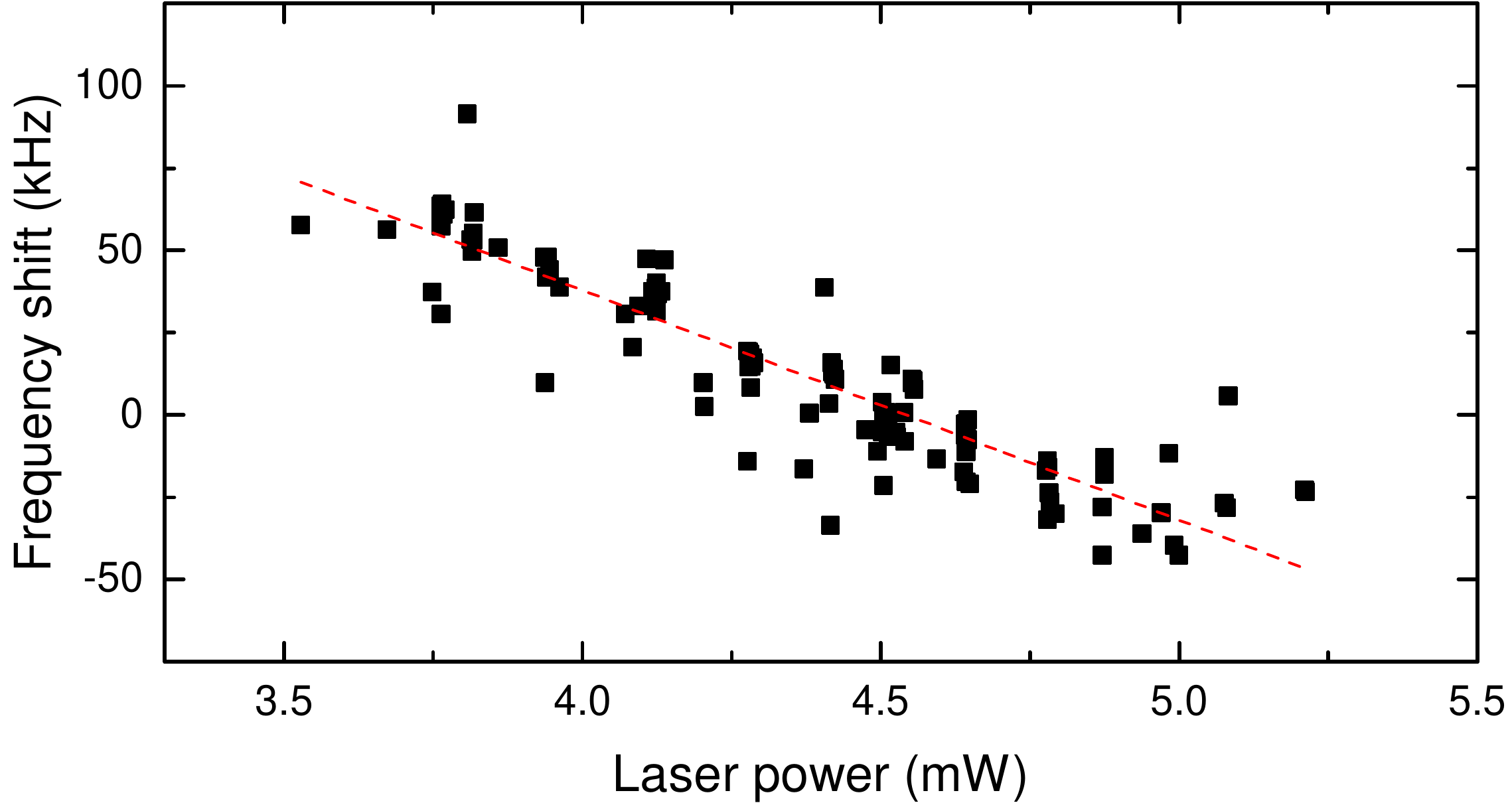}
\caption{Frequency shift of the reference laser relative to optimal laser power recorded versus laser power in the modulation transfer spectroscopy setup. \change{The error bar for each data point is well within the symbol representing it.} For small laser power deviations a linear dependence of 91 kHz per mW is extracted (dashed line). \label{pow_sens}}
\end{figure}

In order to study the influence of laser power fluctuations on the laser frequency we vary the optical power going into the MTS setup while recording both laser power and beat frequency. For relative changes below 10\,\% the laser frequency responds to a good approximation linearly to a change in laser power (see Fig.~\ref{pow_sens}). \change{Choosing the laser power level for the MTS is a tradeoff between a larger error signal for high power and a narrower linewidth for lower power. The optimum power level is defined by a maximum in the error signal slope. After determining this} optimum power level, a sensitivity for absolute laser frequency shift per unit optical power shift \change{was} extracted. \change{With the given laser and rf power} we obtain a sensitivity of 91\,kHz/mW. We observe laser power fluctuations up to $\pm 8\%$ or 0.35 mW.  The resulting fluctuations in the zero-crossing are the main source of residual frequency fluctuations in the reference laser. \change{For more demanding applications, this effect could be further suppressed by active laser power stabilization.}

\change{The linewidth of the reference laser was estimated through comparison with a stable optical reference. For this, we stabilized the optical frequency comb to the optical reference \cite{scharnhorst2015high} and recorded the spectral distribution of the beat note between the reference laser and the frequency comb with a spectrum analyzer.} This signal is a convolution of the $\sim$\,100\,kHz broad comb tooth with the reference laser line. \change{From a Gaussian fit to the signal we derive an upper bound for the linewidth of the reference laser of 400\,kHz.}

\begin{figure}[htb]
\centering\includegraphics[width=0.45\textwidth]{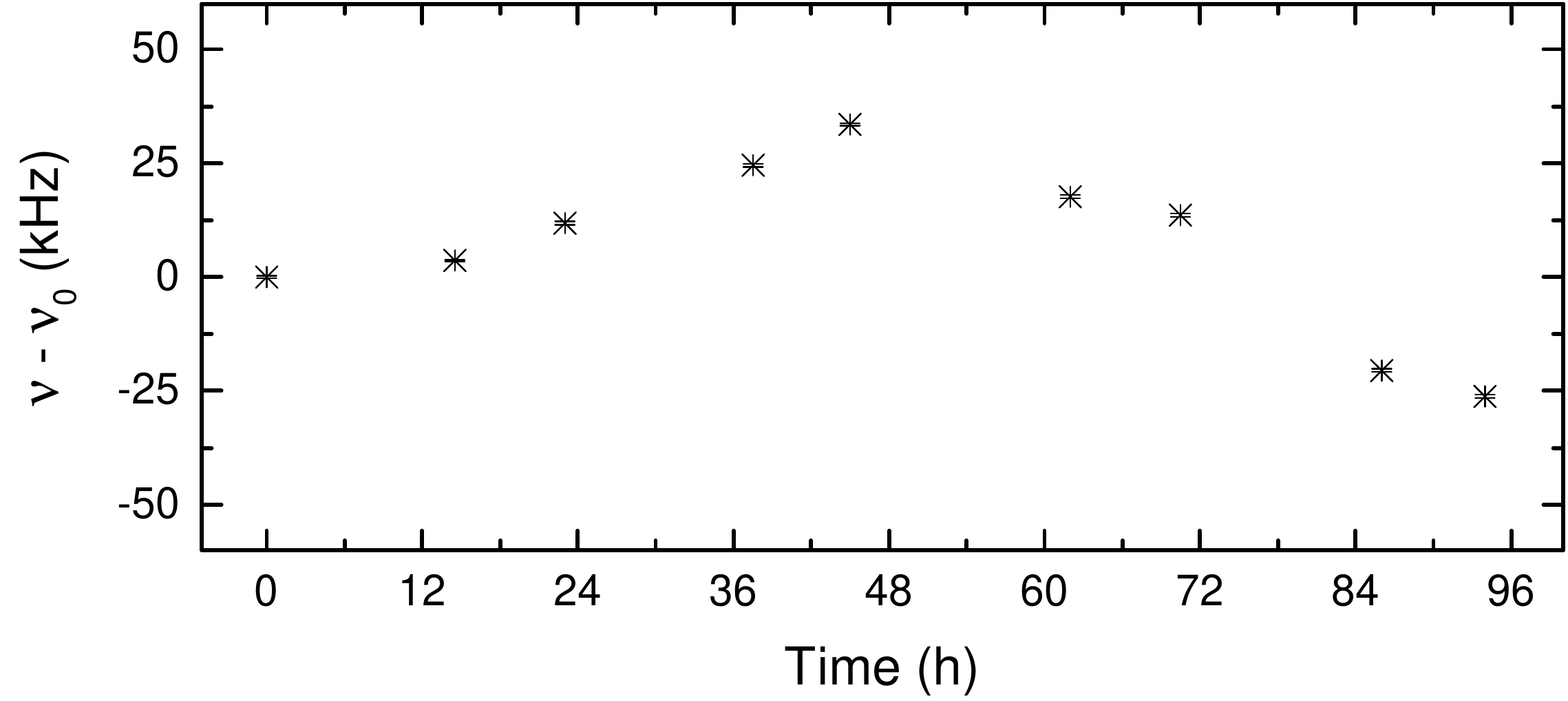}
\caption{Reference laser frequency measurements relative to the absolute frequency $\nu_0=384.229\,240\,061$\,THz of the first data point. Each data point is averaged over 30\,seconds. \change{The error bars for each measurement are well within the size of the symbol representing it.} The offset of 1.6\,MHz from the literature value \cite{steckrubidium85} arises from a finite magnetic field, a pressure shift and the asymmetric error signal. \label{rb_drift}}
\end{figure}

The long-term frequency stability is evaluated by several comparisons against an active hydrogen maser frequency standard with the help of the frequency comb over the course of 4 days. Fig.\,\ref{rb_drift} shows the frequency deviation of the reference laser for each measurement relative to the first measurement. In a time interval of 4~days the change in frequency is well within a 100\,kHz band, the RMS deviation being 20\,kHz. At a wavelength of 780\,nm this corresponds to a fractional instability of the absolute laser frequency of $5\times 10^{-11}$.

\subsection{Transfer lock} \label{sec:cav}
A schematic overview of the cavity transfer lock is shown in the upper right-hand box in Fig.~\ref{setup}. To sustain a transfer lock bridging 100\,nm in the optical/infrared range, care has to be taken to suppress differential dispersion inside the cavity arising from temperature, pressure and relative humidity fluctuations in the air. To that end, evacuated cavities are commonly employed \cite{hough1984dye,helmcke1987new,zhao1998computer}.

\change{We employ a cavity spacer made from a 25\,mm diameter Invar rod with a through hole. Together with a ring piezo attached to one end of the rod, the total cavity length amounts to 100\,mm.} Two curved mirrors with $R=250$\,mm are glued to both ends of the spacer, yielding a symmetric cavity with a free spectral range of about 1500\,MHz. The length of the cavity was chosen such that higher order modes do not coincide with PDH modulation signals. \change{This approach avoids distortion of the main signal from beam pointing fluctuations that change the optical power in these modes \cite{amairi2013reducing}.}

The broadband high-reflective coating of the cavity mirrors reflects more than 99\,\% of the light in the wavelength range between $775\dots 890$\,nm. For the employed wavelengths of 780 and 882\,nm, we measure a finesse of about 300 and 400, respectively. The spacer is placed in a small vacuum chamber with vibration insulated mounting. An attached 2\,l/s ion pump maintains a vacuum of $\sim$\,$10^{-8}$\,mbar. 

The electronic sideband locking technique \cite{thorpe2008laser} based on PDH stabilization is used to stabilize the length of the cavity to the 780\,nm reference laser. The basic idea is to create a sideband with a tunable offset to the carrier laser frequency. This sideband itself exhibits phase modulation sidebands for PDH stabilization. If the variable offset spans a range of half an FSR, the cavity can be locked to one or the other of the offset sidebands. \change{This allows the cavity to be locked to the reference laser for any desired spectroscopy laser frequency.}
The electronic signal necessary to create the above mentioned sidebands is a phase modulated rf signal with variable carrier frequency up to FSR/2, in our case 750\,MHz. 
To create the phase modulated rf signal (see upper right box in Fig.\,\ref{setup}), a VCO (Mini-Circuits ZX95-800C) at $798\dots 803$\,MHz is locked to a fixed-frequency PLL at 800\,MHz, referenced to a maser-stabilized 10\,MHz source. \change{The control input of the VCO used to stabilize the carrier frequency is modulated via a bias tee with $\subsc{\omega}{PDH} \simeq 9.3$\,MHz, thus phase modulating the VCO output at the PDH frequency.} Mixing the VCO output with a signal generator at $810\dots 1600$\,MHz creates sum and difference frequencies. A low-pass filter with a cutoff at 800\,MHz \change{is used} to remove the sum frequency. \change{The difference frequency $\subsc{\omega}{var}$ can be varied between $10\dots 800$\,MHz, while $\subsc{\omega}{PDH}$ remains fixed to $\sim$\,9.3\,MHz.} This signal is applied to a fiber-coupled EOM (Photline NIR-MPX800-LN-05) with an rf bandwidth of 5\,GHz.\par
\begin{figure}[htb]
\centering\includegraphics[width=0.45\textwidth]{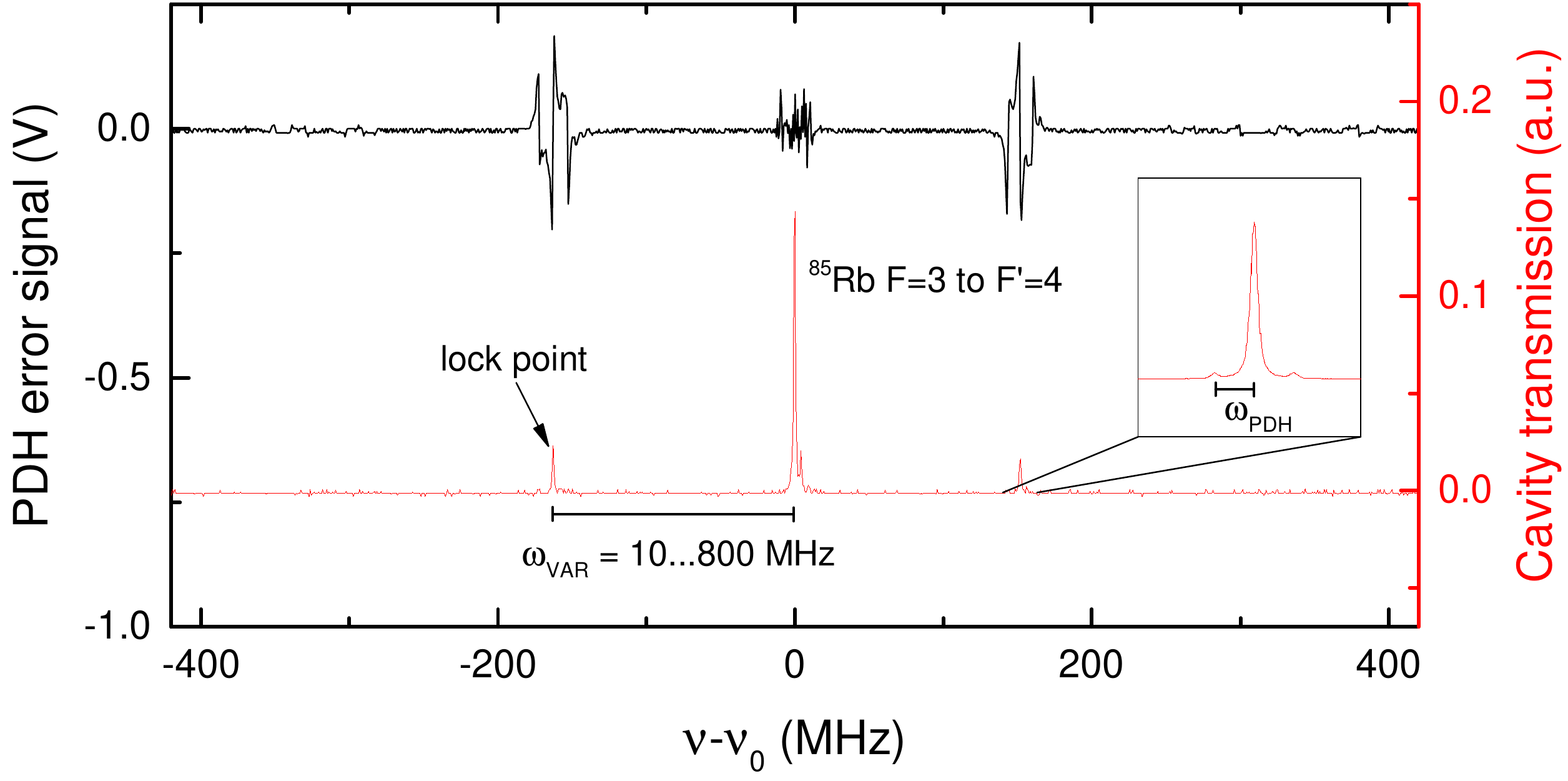}
\caption{Photo diode signal of the transfer cavity transmission around the transition frequency $\nu_0$ for F=3 to F'=4 in $^{85}$Rb  (red) and corresponding PDH error signal (black). 
\change{The inset shows a zoom of the sideband at $\subsc{\omega}{var}$, showing the PDH sidebands at $\subsc{\omega}{var}\pm\subsc{\omega}{PDH}$.} The slopes of the error signals at $\pm\subsc{\omega}{var}$ have opposite signs. \label{SBL}}
\end{figure}

  
Fig.\,\ref{SBL} shows a trace of the error signal obtained in our setup together with the corresponding cavity transmission \change{recorded by a photo diode}. \change{Both sidebands at $\pm \subsc{\omega}{var}$ from the carrier can be used for locking the cavity, thus covering one free spectral range of 1500\,MHz.}

The feedback loop stabilizing the cavity length to the reference laser is electronically bandwidth limited to 150\,Hz. Therefore, frequency noise on the reference laser with Fourier frequencies above 150\,Hz does not affect the cavity length. This allows us to transfer the mean frequency stability of the reference laser without being limited by its high-frequency noise.

\subsection{Spectroscopy laser} \label{sec:spec}

A Ti:Sapphire ring laser (TekhnoScan TIS-SF-07) is used as a light source at 882 nm. It is optically pumped with 10\,watts at 532\,nm by a diode pumped solid state laser (Verdi G10). \change{On the order of 20\,mW of the total output power of $\sim$\,500\,mW is used for locking the laser to the transfer cavity. The remaining power is frequency doubled in an external second harmonic generation stage using a temperature-tuned periodically-poled KTP crystal. An output power of 200 mW at 441 nm is routinely achieved.} 

The beam towards the transfer cavity is phase modulated at 91\,MHz in a resonant-circuit EOM (TimeBase EOM-320IR). A PDH setup generates an error signal from the cavity-reflected beam. The signal is modified by a PI controller which feeds back to two feedback channels acting upon different piezo actuators inside the \change{laser} resonator. \change{The fast channel drives a fast piezo actuator with low stroke. A small mirror mounted on this actuator compensates} high-frequency fluctuations with low modulation index. The resonance frequency of this feedback channel is above 50\,kHz. \change{The slow feedback channel drives three large piezo ring actuators with mounted cavity mirrors.} This channel compensates low-frequency fluctuations up to several 100\,Hz with high amplitude.

\subsection*{Performance of the spectroscopy laser} \label{sec:perf_spec}
\begin{figure}[htb]
\centering\includegraphics[width=0.45\textwidth]{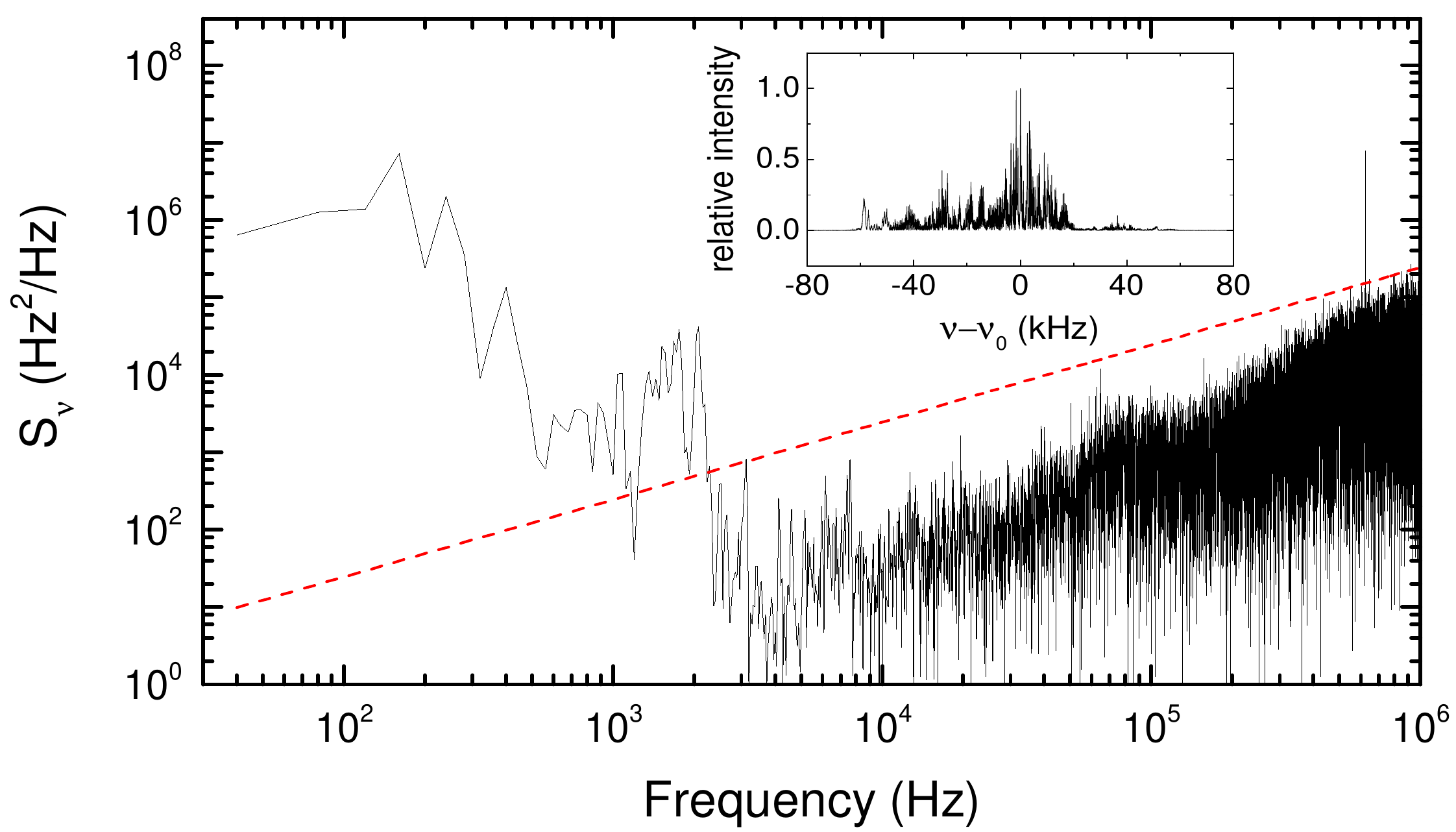}
\caption{Frequency noise spectral density $S_\nu$ of the spectroscopy laser in transfer locked state (solid line). The $\beta$-separation line (dashed line) separates the areas of high and low modulation index. The inset shows a 50\,ms snapshot of the laser frequency compared to another laser with $\sim$\,1\,Hz linewidth. \label{linewidth}}
\end{figure}

The linewidth of the locked spectroscopy laser is characterized by comparison with an ultra-stable laser at 1542\,nm, locked to a high-finesse passive cavity resulting in a linewidth of $\sim$\,1\,Hz. \change{For both lasers a beat note with the frequency comb and additionally the carrier envelope offset is recorded using a 200 MS/s digital oscilloscope which digitizes $10^7$ data points within 50 ms. A numerical evaluation according to \cite{telle2002kerr} yields the frequency difference over time between our spectroscopy laser and the ultra-stable laser. As the frequency variations of the ultra-stable laser can be neglected on that time scale, fluctuations in the frequency difference can be assigned to absolute frequency fluctuations of the spectroscopy laser.} The inset of Fig.~\ref{linewidth} shows the distribution \change{of the frequency fluctuations of the spectroscopy laser} around the mean value. \change{As the spectral distribution is not Gaussian}, an estimate for the linewidth is inferred from the frequency noise spectral density. In Ref.~\cite{di2010simple} a method is presented to extract a laser linewidth from arbitrary spectral noise distributions. 
The frequency noise spectrum can be divided into two regions. One with a high noise level compared to the Fourier frequency, thus exerting a modulation index $\beta > 1$, and one with a low noise level compared to the Fourier frequency, yielding a low modulation index. The first region contributes to the linewidth, while the latter mostly affects the wings of the laser line. These regions are separated by the so-called $\beta$-separation line. \change{The full width at half maximum (FWHM) of the laser linewidth is extracted by integrating the frequency noise spectral density, $S_\nu$, up to the Fourier frequency where it intersects the $\beta$-separation line. In our case this is at about 2.2 kHz (see Fig.\,\ref{linewidth}). The lower cut-off frequency is given by the observation time of 50 ms and corresponds to 40 Hz. For the integration we most likely overestimate $S_\nu$ by assuming a constant value between 0-40 Hz.} Applying this method we derive a FWHM of 55\,kHz. The RMS deviation of the laser frequency distribution in the inset of Fig.\,\ref{linewidth} is 23\,kHz. \change{We expect that frequency doubling in an external cavity doubles the linewidth without introducing significant additional noise \cite{liu2007narrow}. Frequency doubling is performed with a PPKTP crystal in a bow-tie external doubling cavity. We routinely achieve an output power of 200\,mW at 441\,nm with 500\,mW power at the fundamental frequency.}

\begin{figure}[htb]
\centering\includegraphics[width=0.45\textwidth]{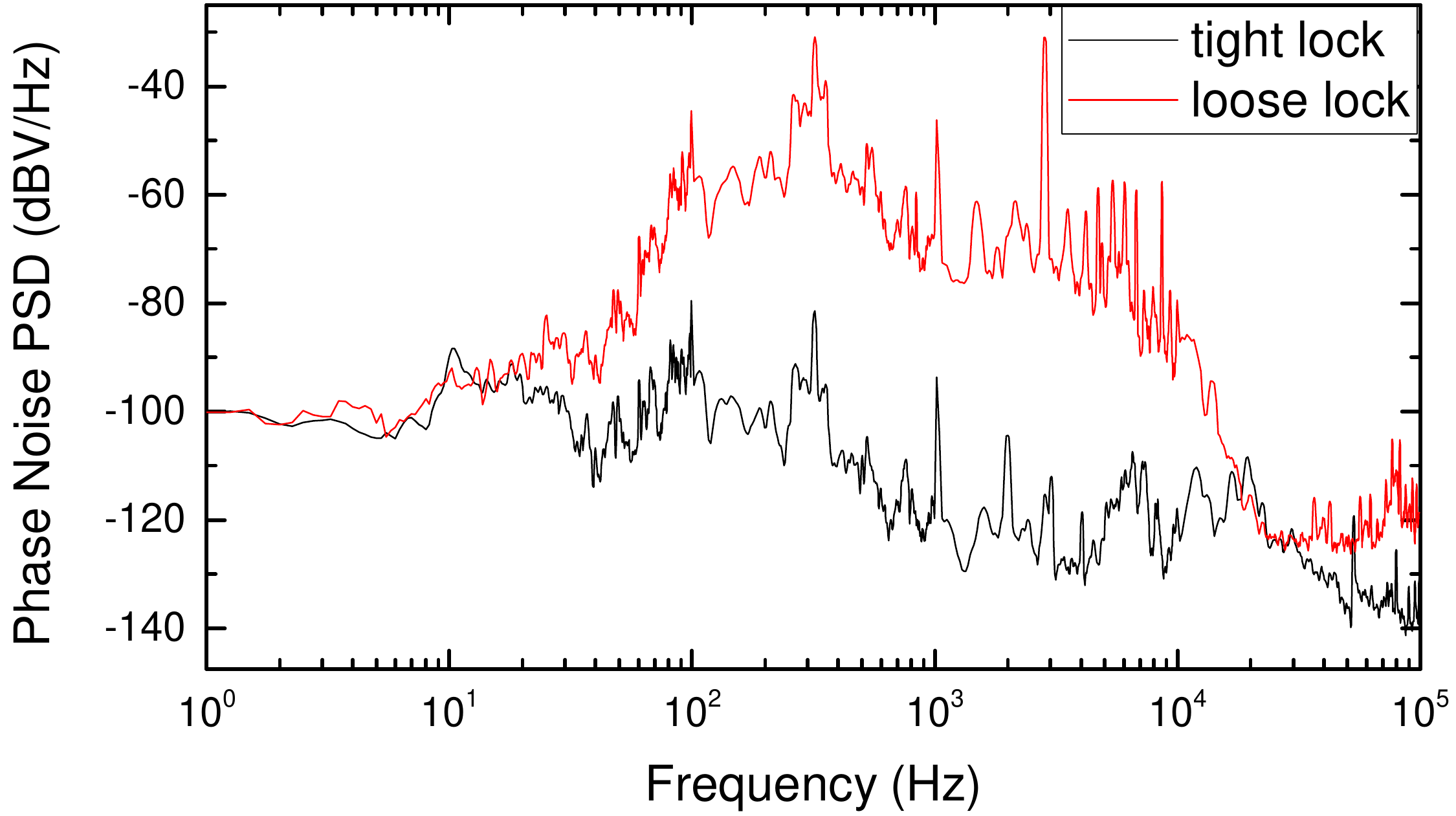}
\caption{Phase noise power spectral density of the Ti:Sapphire laser in tight lock (black) and loose lock (red). A noise suppression of up to 50\,dB is established by the cavity lock. We infer a feedback bandwidth of around 18\,kHz for the fast feedback channel from the crossing point of the two curves. The difference in PSD above 18\,kHz between both curves \change{arises from excursions of the laser frequency beyond the linear region of the PDH error signal, leading to high-frequency harmonics of
low-frequency noise components.} \label{noise}}
\end{figure}

\FloatBarrier
We characterize \change{the laser frequency} stabilization by comparing the power spectral densities (PSD) of the in-loop error signal for two cases: a tight lock with optimized parameters and a loose lock where the stabilization is just sufficient to keep the laser frequency on the central slope of the PDH error signal. That way, we obtain a nearly linear voltage response for frequency fluctuations. Fig.\,\ref{noise} shows PSD spectra up to 100 kHz recorded with a fast Fourier transform device. It can be seen that the feedback loop suppresses noise within a bandwidth of about 18\,kHz.

With the transfer lock engaged, the frequency instability of the spectroscopy laser is measured the same way as before for the reference laser. The wavelength of the spectroscopy laser is tuned to 871\,nm where an output port of a frequency comb is available. The beat frequency between spectroscopy laser and frequency comb is monitored over several days. 
\begin{figure}[htb]
\centering\includegraphics[width=0.45\textwidth]{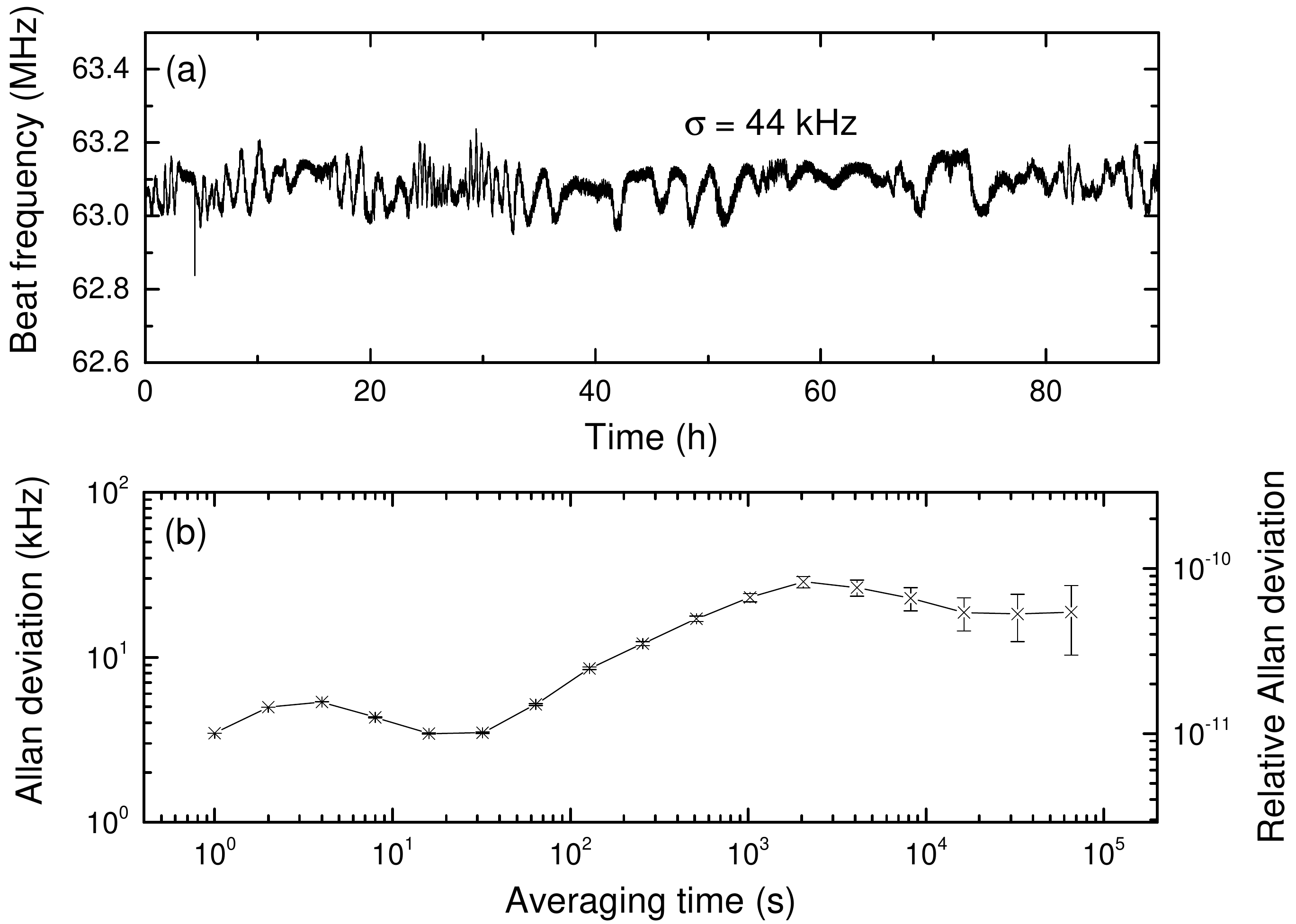}
\caption{(a) Beat frequency of the transfer locked spectroscopy laser with a frequency comb at 871\,nm. (b) Absolute and relative overlapping Allan deviation of the transfer locked spectroscopy laser up to averaging times of $6.5\times 10^{4}$\,s. \label{drift_tisa}}
\end{figure}
Fig.\,\ref{drift_tisa} (a) shows a frequency trace taken over 90~hours. A linear fit to the data suggests a frequency drift of $(0.28 \pm 0.30)$\,kHz/h. The RMS frequency deviation is 44\,kHz. More information on the stability can be extracted from the Allan deviation in Fig.\,\ref{drift_tisa} (b). We achieve a frequency instability below $10^{-10}$ for averaging times up to 18\,h. At that point the Allan deviation flattens out, indicating no linear drift on this time scale. \change{The oscillations in the laser frequency visible in Fig.\,\ref{drift_tisa} (a) (also visible as local maxima in the Allan deviation in Fig.\,\ref{drift_tisa} (b)), are most likely of thermal origin. The two characteristic times of about 4 s and 2000 s can be traced back to the control loop for laser diode temperature and room temperature, respectively. Oscillations in temperature affect the laser power and any spurious etalon fringes between optical elements. This leads to frequency fluctuations on these time scales which are, however, within the 50\,kHz tolerance required for the intended application.}\par 
We expect that the lower limit in Allan deviation is given by the instability of the reference laser. The stability of the spectroscopy laser is further deteriorated by offset fluctuations in the PDH error signals due to residual amplitude modulation.

\section{Conclusion}
\change{We have presented a laser setup that is capable of providing a gapless output range spanning over 100\,nm in the infrared spectrum. We have demonstrated that the fast linewidth and long-term frequency drifts are below 50\,kHz over several days.} 
These values were verified by measurements with a referenced optical frequency comb. We show that the spectral impurity of the reference laser does not limit the spectroscopy laser linewidth due to low-pass filtering of frequency noise of the transfer cavity. The transfer locking technique demonstrated here can be extended to cover a broad range of wavelengths in the optical and infrared, limited only by the cavity mirror coating. \change{To our knowledge this is the best combined performance in terms of linewidth and long-term stability for a transfer lock using a broad alkali D-line as reference transition. Better results can be achieved on narrower atomic transitions but require an intrinsically narrower linewidth of the reference laser \cite{Bridge:16}.}\par 
Current limitations in our frequency stabilization setup arise mainly from the reference laser. A simple improvement would be a laser power stabilization of the reference laser and a temperature stabilization of the EOM used for the modulation transfer spectroscopy to reduce time-varying RAM. Furthermore, \change{better thermal isolation} of the rubidium vapor cell would further improve the stability of the reference. To narrow the short term linewidth of the spectroscopy laser, a higher feedback bandwidth is necessary. An intracavity EOM or external AOM \cite{haubrich1996modified,boyd1991frequency} could be used for this purpose. Additionally, the cavity mirror coating should then be tailored in such a way, that the finesse is substantially higher for the spectroscopy laser while keeping a value on the order of 100 for the reference laser.\par 
With this setup we anticipate to resolve the line of the fine structure transition $1s^{2}2s^{2}2p\ ^{2}P_{1/2} -\ ^{2}P_{3/2}$ at 441\,nm in Coulomb-crystallized Ar$^{13+}$ ions in a Paul trap to within 100\,kHz.

\section*{Acknowledgments}
We acknowledge support from DFG through QUEST. I.D.L. acknowledges a
fellowship from the Alexander von Humboldt Foundation. This work was
funded by PTB.


\begin{thebibliography}{10}
\providecommand{\url}[1]{\texttt{#1}}
\providecommand{\urlprefix}{URL }

\bibitem{berengut2010enhanced}
J.~Berengut, V.~Dzuba, and V.~Flambaum, Phys. Rev. Lett. \textbf{105}, 120801
  (2010).

\bibitem{derevianko_highly_2012}
A.~Derevianko, V.~Dzuba, and V.~Flambaum, Physical review letters \textbf{109},
  180801 (2012).

\bibitem{dzuba_optical_2015}
V.~Dzuba, V.~Flambaum, and H.~Katori, Physical Review A \textbf{91}, 022119
  (2015).

\bibitem{berengut2012highly}
J.~Berengut, V.~Dzuba, V.~Flambaum, and A.~Ong, Phys. Rev. A \textbf{86},
  022517 (2012).

\bibitem{safronova2014highly}
M.~Safronova, V.~Dzuba, V.~Flambaum, U.~Safronova, S.~Porsev, and M.~Kozlov,
  Phys. Rev. Lett. \textbf{113}, 030801 (2014).

\bibitem{mackel2011laser}
V.~M{\"a}ckel, R.~Klawitter, G.~Brenner, J.~C. L{\'o}pez-Urrutia, and
  J.~Ullrich, Phys. Rev. Lett. \textbf{107}, 143002 (2011).

\bibitem{schmoger2015coulomb}
L.~Schm{\"o}ger, O.~O. Versolato, M.~Schwarz, M.~Kohnen, A.~Windberger,
  B.~Piest, S.~Feuchtenbeiner, J.~Pedregosa-Gutierrez, T.~Leopold, P.~Micke,
  A.~K. Hansen, T.~M. Baumann, M.~Drewsen, J.~Ullrich, P.~O. Schmidt, and J.~R.
  Crespo L\'opez-Urrutia, Science \textbf{347}, 1233 (2015).

\bibitem{young_visible_1999}
B.~C. Young, F.~C. Cruz, W.~M. Itano, and J.~C. Bergquist, Phys. Rev. Lett.
  \textbf{82}, 3799--3802 (1999).

\bibitem{kessler_sub-40-mhz-linewidth_2012}
T.~Kessler, C.~Hagemann, C.~Grebing, T.~Legero, U.~Sterr, F.~Riehle, M.~J.
  Martin, L.~Chen, and J.~Ye, Nature Photon. \textbf{6}, 687--692 (2012).

\bibitem{hafner_8_2015}
S.~H{\"a}fner, S.~Falke, C.~Grebing, S.~Vogt, T.~Legero, M.~Merimaa, C.~Lisdat,
  and U.~Sterr, Opt. Lett. \textbf{40}, 2112 (2015).

\bibitem{benkler_robust_2013}
E.~Benkler, F.~Rohde, and H.~R. Telle, Opt. Lett. \textbf{38}, 555--557 (2013).

\bibitem{inaba_spectroscopy_2013}
H.~Inaba, K.~Hosaka, M.~Yasuda, Y.~Nakajima, K.~Iwakuni, D.~Akamatsu, S.~Okubo,
  T.~Kohno, A.~Onae, and F.-L. Hong, Opt. Express \textbf{21}, 7891--7896
  (2013).

\bibitem{rohde_phase-predictable_2014}
F.~Rohde, E.~Benkler, T.~Puppe, R.~Unterreitmayer, A.~Zach, and H.~R. Telle,
  Opt. Lett. \textbf{39}, 4080--4083 (2014).

\bibitem{nicolodi2014spectral}
D.~Nicolodi, B.~Argence, W.~Zhang, R.~Le~Targat, G.~Santarelli, and Y.~Le~Coq,
  Nature Photon. \textbf{8}, 219--223 (2014).

\bibitem{scharnhorst2015high}
N.~Scharnhorst, J.~B. W\"{u}bbena, S.~Hannig, K.~Jakobsen, J.~Kramer, I.~D.
  Leroux, and P.~O. Schmidt, Opt. Express \textbf{23}, 19771--19776 (2015).

\bibitem{wieman_doppler-free_1976}
C.~Wieman and T.~W. H{\"a}nsch, Phys. Rev. Lett. \textbf{36}, 1170--1173
  (1976).

\bibitem{corwin_frequency-stabilized_1998}
K.~L. Corwin, Z.-T. Lu, C.~F. Hand, R.~J. Epstein, and C.~E. Wieman, Appl. Opt.
  \textbf{37}, 3295--3298 (1998).

\bibitem{ye_ultrasensitive_1998}
J.~Ye, L.-S. Ma, and J.~L. Hall, JOSA B \textbf{15}, 6--15 (1998).

\bibitem{shirley1982modulation}
J.~H. Shirley, Opt. Lett. \textbf{7}, 537 (1982).

\bibitem{bjorklund_frequency_1983}
G.~C. Bjorklund, M.~D. Levenson, W.~Lenth, and C.~Ortiz, Appl. Phys. B
  \textbf{32}, 145--152 (1983).

\bibitem{mccarron2008modulation}
D.~J. McCarron, S.~A. King, and S.~L. Cornish, Meas. Sci. Technol. \textbf{19},
  105601 (2008).

\bibitem{riedle1994stabilization}
E.~Riedle, S.~H. Ashworth, J.~T. Farrell~Jr., and D.~J. Nesbitt, Rev. Sci.
  Instrum. \textbf{65}, 42 (1994).

\bibitem{vogt2011demonstration}
S.~Vogt, C.~Lisdat, T.~Legero, U.~Sterr, I.~Ernsting, A.~Nevsky, and
  S.~Schiller, Appl. Phys. B \textbf{104}, 741--745 (2011).

\bibitem{chen2014compact}
Q.-F. Chen, A.~Nevsky, M.~Cardace, S.~Schiller, T.~Legero, S.~H{\"a}fner,
  A.~Uhde, and U.~Sterr, Rev. Sci. Instrum. \textbf{85}, 113107 (2014).

\bibitem{leibrandt2011field}
D.~R. Leibrandt, M.~J. Thorpe, J.~C. Bergquist, and T.~Rosenband, Opt. Express
  \textbf{19}, 10278--10286 (2011).

\bibitem{leibrandt2011spherical}
D.~R. Leibrandt, M.~J. Thorpe, M.~Notcutt, R.~E. Drullinger, T.~Rosenband, and
  J.~C. Bergquist, Opt. Express \textbf{19}, 3471--3482 (2011).

\bibitem{dube2009narrow}
P.~Dub{\'e}, A.~Madej, J.~Bernard, L.~Marmet, and A.~Shiner, Appl. Phys. B
  \textbf{95}, 43--54 (2009).

\bibitem{matthey2015compact}
R.~Matthey, F.~Gruet, S.~Schilt, and G.~Mileti, Optics letters \textbf{40},
  2576--2579 (2015).

\bibitem{bohlouli-zanjani_optical_2006}
P.~Bohlouli-Zanjani, K.~Afrousheh, and J.~D.~D. Martin, Rev. Sci. Instrum.
  \textbf{77}, 093105 (2006).

\bibitem{uetake2009frequency}
S.~Uetake, K.~Matsubara, H.~Ito, K.~Hayasaka, and M.~Hosokawa, Appl. Phys. B
  \textbf{97}, 413 (2009).

\bibitem{rohde2010diode}
F.~Rohde, M.~Almendros, C.~Schuck, J.~Huwer, M.~Hennrich, and J.~Eschner, J.
  Phys. B \textbf{43}, 115401 (2010).

\bibitem{albrecht_laser_2012}
S.~Albrecht, S.~Altenburg, C.~Siegel, N.~Herschbach, and G.~Birkl, Appl. Phys.
  B \textbf{107}, 1069--1074 (2012).

\bibitem{yin_narrow-linewidth_2015}
Y.~Yin, Y.~Xia, X.~Li, X.~Yang, S.~Xu, and J.~Yin, Appl. Phys. Express
  \textbf{8}, 092701 (2015).

\bibitem{Bridge:16}
E.~M. Bridge, N.~C. Keegan, A.~D. Bounds, D.~Boddy, D.~P. Sadler, and M.~P.~A.
  Jones, Opt. Express \textbf{24}, 2281--2292 (2016).

\bibitem{black_introduction_2001}
E.~D. Black, Am. J. Phys. \textbf{69}, 79 (2001).

\bibitem{steckrubidium85}
D.~A. Steck, \emph{Rubidium 85 {D} Line Data (revision 2.1.4, 2010)}.

\bibitem{steckrubidium87}
D.~A. Steck, \emph{Rubidium 87 {D} Line Data} (2001).

\bibitem{jaatinen1998possible}
E.~Jaatinen and J.~M. Chartier, Metrologia \textbf{35}, 75 (1998).

\bibitem{MartinezdeEscobar:15}
Y.~N.~M. de~Escobar, S.~P. \'{A}lvarez, S.~Coop, T.~Vanderbruggen, K.~T.
  Kaczmarek, and M.~W. Mitchell, Opt. Lett. \textbf{40}, 4731 (2015).

\bibitem{PhysRev.124.800}
M.~Arditi and T.~R. Carver, Phys. Rev. \textbf{124}, 800 (1961).

\bibitem{hough1984dye}
J.~Hough, D.~Hils, M.~D. Rayman, L.-S. Ma, L.~Hollberg, and J.~L. Hall, Appl.
  Phys. B \textbf{33}, 179 (1984).

\bibitem{helmcke1987new}
J.~Helmcke, J.~J. Snyder, A.~Morinaga, F.~Mensing, and M.~Gl{\"a}ser, Appl.
  Phys. B \textbf{43}, 85 (1987).

\bibitem{zhao1998computer}
W.~Z. Zhao, J.~E. Simsarian, L.~A. Orozco, and G.~D. Sprouse, Rev. Sci.
  Instrum. \textbf{69}, 3737 (1998).

\bibitem{amairi2013reducing}
S.~Amairi, T.~Legero, T.~Kessler, U.~Sterr, J.~B. W{\"u}bbena, O.~Mandel, and
  P.~O. Schmidt, Appl. Phys. B \textbf{113}, 233--242 (2013).

\bibitem{thorpe2008laser}
J.~Thorpe, K.~Numata, and J.~Livas, Opt. Express \textbf{16}, 15980--15990
  (2008).

\bibitem{telle2002kerr}
H.~R. Telle, B.~Lipphardt, and J.~Stenger, Appl. Phys. B \textbf{74}, 1 (2002).

\bibitem{di2010simple}
G.~Di~Domenico, S.~Schilt, and P.~Thomann, Appl. Opt. \textbf{49}, 4801 (2010).

\bibitem{liu2007narrow}
T.~Liu, Y.~Wang, R.~Dumke, A.~Stejskal, Y.~Zhao, J.~Zhang, Z.~Lu, L.~Wang,
  T.~Becker, and H.~Walther, Applied Physics B \textbf{87}, 227--232 (2007).

\bibitem{haubrich1996modified}
D.~Haubrich and R.~Wynands, Opt. Commun. \textbf{123}, 558--562 (1996).

\bibitem{boyd1991frequency}
T.~L. Boyd and H.~Kimble, Opt. Lett. \textbf{16}, 808--810 (1991).

\end{thebibliography}

\end{document}